\documentstyle[twoside,psfig,mpla1]{article}

\def\b{\begin{equation}}
\def\e{\end{equation}}
\def\br{\begin{eqnarray}}
\def\er{\end{eqnarray}}
\def\pa{\partial}
\def\l{\left}
\def\r{\right}

\begin{document}
\setlength{\textheight}{7.7truein}  


\runninghead{Method of complex paths and general covariance of Hawking 
radiation}{Shankaranarayanan, Srinivasan, Padmanabhan}

\normalsize
\textlineskip
\thispagestyle{empty}
\setcounter{page}{1}
\copyrightheading{}                     
\vspace*{0.88truein}
\fpage{1}


\centerline{Method of complex paths and general covariance of Hawking
radiation}
\vspace*{0.37truein}
\centerline{\footnotesize S.~Shankaranarayanan\footnote{Electronic
address:~shanki@iucaa.ernet.in}}
\centerline{\footnotesize K.~Srinivasan}
\centerline{\footnotesize T.~Padmanabhan\footnote{Electronic
address:~paddy@iucaa.ernet.in}}
\baselineskip=12pt
\centerline{\footnotesize\it IUCAA, Post Bag 4, Ganeshkhind}
\baselineskip=10pt
\centerline{\footnotesize\it Pune 411 007, INDIA.}
\vspace*{0.225truein}

\publisher{(received date)}{(revised date)}

\vspace*{0.21truein}

\abstracts{ We apply the technique of complex paths to obtain Hawking 
radiation in different coordinate representations of the Schwarzschild
space-time. The coordinate representations we consider do not possess a
singularity at the horizon unlike the standard Schwarzschild coordinate.
However, the event horizon manifests itself as a singularity in the expression
for the semi-classical action. This singularity is regularized by using
the method of complex paths and we find that Hawking radiation is recovered in 
these coordinates indicating the covariance of Hawking radiation. This 
also shows that there is no correspondence between the particles detected by 
the model detector and the particle spectrum obtained by the quantum field 
theoretic analysis -- a result known in other contexts as well.}{}{}

\vspace*{1.5truecm}\textlineskip      

	Quantum field theory(QFT) requires the notion of a time-like killing 
vector field to define particles. When curvilinear coordinate transformations 
are allowed in flat space-time, there exists other possible killing vector 
fields which are time-like in part of the manifold. Particle states -- in 
particular vacuum state -- can be defined with respect to these killing vector 
fields. In general, these definitions will not be equivalent and the Minkowski 
vacuum will appear to be a many particle states according to the new 
definition. Thus, the presence of, say, Rindler quanta in the Minkowski vacuum 
arises because there 
is more than one way of defining positive modes in a given space-time, even 
though the space-time itself is static. On the other hand, particles are 
created in a time dependent metric because the natural definition of positive 
frequency modes are different at two different times. This is interpreted as 
the production of particles corresponding to quantum field by the changing 
geometry of spacetime.
\par 
	  The most spectacular prediction of quantum field theory in curved 
space-time is undoubtedly the Hawking radiation from a black hole. 
Hawking\cite{hawking75} showed that QFT in the background of a body collapsing 
to a black hole predicts a radiation of particles at late times with 
characteristic thermal spectrum at temperature equal to $1/8\pi M $. But, as 
mentioned in the previous paragraph, the concept of a particle in quantum field 
theory is not generally covariant and depends on the coordinates chosen to 
describe the particular space-time. It is therefore of interest to study the 
Hawking effect in other coordinate representations. The method used by Hawking 
in his analysis requires the knowledge of the wave modes of the quantum field 
in the standard Schwarzschild coordinate(SSC). In attempting to study 
Hawking effect in other coordinate systems, one runs into problem of 
identifying the wave modes and calculating Bogoliubov coefficients to identify 
the spectrum of the radiation which is intractable. Hence, it is necessary to 
have a method which does not use wave modes to calculate the emission 
spectrum. 
\par
	Hartle and Hawking\cite{hawking76} obtained particle 
production in the standard black hole space-times using semiclassical 
analysis which does not require wave modes. In this method, the semiclassical 
propagator for a scalar field propagating in the maximally extended Kruskal 
manifold is analytically continued in the time variable $t$ to complex values. 
This analytic continuation gives the result that the probability of emission 
of particles from the past horizon is not the same as the probability of 
absorption into the future horizon. The ratio between these probabilities is 
of the form
\b
P[{\textrm{emission}}] = P[{\textrm{absorption}}] e^{-\beta E},
\e
where $E$ is the energy of the particles and $\beta^{-1}=1/8\pi M$ is the 
standard Hawking temperature. The above relation is interpreted to be 
equivalent to a thermal distribution of particles in analogy with that 
observed in any system interacting with black body radiation. In this 
analysis, the probability amplitude for the emission/absorption is calculated 
by identifying a particular path for both these processes in the fully 
extended Kruskal manifold. 
\par
Unfortunately the Kruskal extension is of vital importance in obtaining the 
thermal spectrum in this analysis, and hence, it cannot be adapted for other 
coordinate systems. Recently, Srinivasan and Padmanabhan\cite{kt99} (SP,
hereafter) have obtained Hawking radiation {\it without using the Kruskal 
extension}. They have shown that the coordinate singularity present at the 
horizon, in the SSC, manifests itself as a singularity in the expression for 
the semi-classical propagator $K(r_2, t_2; r_1, t_1)$, which is given by 
\b
K(r_2, t_2; r_1, t_1) = N \exp\l[i S_0(r_2, t_2; r_1, t_1)/\hbar \r].
\e
\noindent $S_0$ is the action functional satisfying the classical 
Hamilton-Jacobi equation for a massless particle to propagate from $(t_1, r_1)$ 
to $(t_2, r_2)$ and $N$ is the suitable normalization constant. The authors 
used the method of complex paths (of non-relativistic quantum 
mechanics\cite{landau3}) and modified 
appropriately to produce a prescription that regularizes the singularity in 
the action functional and Hawking radiation was recovered as a consequence. 
The semi-classical propagator obtained in this fashion is an exact propagator 
of the quantum field. The boundary conditions near the horizon clearly shows
that the authors use Unruh vacuum to define particles at future infinity.  
\par
	In this letter, we apply the method of complex paths to two non-static 
coordinate representations of the Schwarzschild space-time and recover Hawking 
radiation. The two coordinates we consider are: Lemaitre coordinate(LC) which 
is a time dependent system and Painleve coordinate(PC) which is a stationary 
system. The line element corresponding to LC is, 
\b
\!\! ds^2\!\! =  d\tau^2\!\! -\!\!\left[\frac{\d 3}{\d 4M}(R\pm\tau)\right]^
{-2/3}\!\!\!\!\!\!\!\!\!\! dR^2\!\! -\!\! \left[{\sqrt{2M}}\frac{3}{2}
\! (R\pm\tau)\right]^{4/3}\!\!\!\!\!\!\!\! d\Omega^2,  
\label{lemaitres}
\e  
where $c = G = 1$ and $d\Omega^2$ is the angular line element. The line 
elements can be modeled as that natural to a freely falling observer whose 
velocity at radial infinity is zero. The time coordinate $\tau$ measures the 
proper time of free falling observers; each observer moves along a line 
$ R = constant$. The lower sign can be represented as particle trajectories 
moving inward to the singularity($r=0$), while the upper sign represents 
that of moving outward from the singularity. 
\par
	The line element corresponding to PC is, 
\b
ds^2\!\! =\! (d\tau_P)^2\!\! -\!\!\l[\! dr\pm\sqrt{\frac{2M}{r}} d\tau_P\r]^2
\!\!\!\! - r^2\! (d\theta^2\!\!+\sin^2\!\theta d\phi^2).
\label{painleves}
\e 
The physical interpretation of these metrics is obtained by comparing them 
with LC. Noting that the constant time slice is a flat Euclidean space for the 
Painleve metric, we transform the coordinate variables of Lemaitre as   
\b
\tau_P = \tau,\ \ r = (2M)^{1/3} \l[3 (R - \tau)/2 \r]^{2/3}\!\!\!
\label{maintrans}
\e
\noindent to go from (\ref{lemaitres}) to (\ref{painleves}). Under these 
transformations, the lower sign in line element (\ref{lemaitres}) leads to 
(\ref{painleves}) with upper sign; repeating the 
same steps starting from upper sign in (\ref{lemaitres}) leads to the other 
line element. 
\par
	Before we proceed with the analysis of quantum fields, it is necessary 
to understand certain important differences between the non static 
representations, LC and PC, with that of SSC. In both these representations, 
LC and PC, metric possesses no coordinate singularity at the horizon while SSC 
possesses a coordinate singularity at the horizon. The SSC is time reversal 
invariant ($t \rightarrow -t$), while the two coordinates, LC and PC, are not. 
In these two coordinates, the time reversal transformation on the metric 
corresponds to a different physical situation. However, the line elements 
corresponding to these different physical situations are related to the SSC by 
a coordinate transformation. The two line elements of Painleve metric can be 
obtained by the coordinate transformation
\b 
\tau_P=t \pm 2\sqrt{2M\,r} \pm 2M\ln\l(\frac{\sqrt{r}-\sqrt{2M}}{\sqrt{r}+
\sqrt{2M}}
\r)
\label{paint1}.
\e
where, $t$ is the time coordinate of SSC (the coordinate transformation 
between LC and SSC can be found in Landau\cite{landau2}). 
\par
	In principle, calculating the probability amplitude for the emission 
and absorption process in these space-times requires identifying a particular 
path for both these processes, as done by Hartle and Hawking\cite{hawking76}, 
in the maximally extended Kruskal manifold which requires 
detailed knowledge of particle trajectories. It is, however, possible to 
analyze these two coordinate systems using the notion of R and T regions, 
introduced by Novikov\cite{novikov64}, which provides an elegant method in 
understanding the global structure of the spacetime, thus making a detailed 
analysis of particle trajectories unnecessary. The general expression for the 
interval in a spherically symmetric field is,
\b
\!\! ds^2\!\! = e^{\nu(x^0, x^1)} (dx^0)^2\! - e^{\lambda(x^0, x^1)} 
(dx^1)^2\! -e^{\mu(x^0, x^1)} d\Omega^2\!\!
\label{sphesymm}.
\e
If at the given event in the general coordinate system, the inequality,
\b
e^{\nu - \lambda} > \l( \frac{\pa\mu}{\pa x^0}/ \frac{\pa\mu}{\pa x^1}\r)^2
\label{maincond}
\e
\noindent is satisfied, then the event is defined as R-region. [If the
above inequality is satisfied at a certain world point then by the virtue of
the continuity ($\exp(\nu - \mu)$, $\pa \mu/\pa x^0$, $\pa \mu/\pa x^1$
cannot be discontinuous) it is satisfied in some neighborhood of this point. 
Thus, the points in the neighborhood of this system of coordinates satisfies 
the above inequality are R-points and a set of them a R-region.] If the 
opposite inequality is satisfied, the event is in a T-region. The definitions 
of R and T regions can be shown to be coordinate invariant. 
\par
	For the LC, we find that the region outside the 
Schwarzschild sphere [for the lower sign in line element (\ref{lemaitres})] 
the inequality $\frac{3}{2} (R - \tau) > 2M$ corresponds to R-region; while the 
region inside the Schwarzschild sphere, where the inequality is opposite to 
the above inequality, is a T-region. In the T-region there is an obvious 
asymmetry in the direction of time flow. For this line element, all motions 
in the T-region($T_-$) are directed towards $r=0$ where the curvature 
invariants are infinite. The upper sign in line element (\ref{lemaitres}) will 
define second type of T-regions --- an expanding region $T_+$. This has 
completely opposite properties and all bodies move away from the singularity 
to the R-regions. For a physically realizable event, the coexistence of $T_+$ 
and $T_-$ regions is impossible. In other words, the T region of the standard 
Schwarzschild is a doubly mapped T region in Lemaitre. 
\par
	In the case of PC, using the transformation relating Lemaitre and 
Painleve in Eq.~(\ref{maintrans}), we find that the inequality $r > 2M$ 
corresponds to R-region for both the line elements in (\ref{painleves}). 
Hence, for both these line elements, R and T regions are the same --- the 
whole of the space-time is doubly mapped.	
\par
	The semi-classical propagator satisfies the relativistic 
Hamilton-Jacobi equation of a (massless/massive) particle moving in the 
particular coordinate system. The solution to the Hamilton Jacobi in the two 
coordinate systems, PC and LC, can be obtained [see Eqs. (12) and (17)]. 
The solution to the Hamilton-Jacobi has the sign ambiguity (of the 
square root) and this is related to the outgoing 
($p_x = \partial S_0/\partial x > 0$) or ingoing 
($\partial S_0/\partial x < 0$) nature of the particle. The momentum
of the particle moving in the $x$-direction is given by $p_x =
\partial S_0/\partial x $. Hence, $p_x > 0$ corresponds to particle
moving away from the horizon (in the case of Schwarzschild
space-time).
\par	 
	 We will now proceed to study QFT in these coordinate systems. We are 
interested in the quantization of a massless scalar field 
$\Phi$ in these two backgrounds. The system we first consider is the lower 
sign in line element (\ref{painleves}) of Painleve. Since all the relevant 
physics is contained in the $(\tau_P,r)$ plane, we set $\Phi = \Psi(\tau_P, r)
Y^m_l(\theta, \phi)$ and concentrate on $\Psi$. The equation satisfied by 
$\Psi$ is,
\br
& &\frac{\pa^2\Psi}{\pa \tau_P^2}-\sqrt{\frac{8M}{r}} \frac{\pa^2\Psi}{\pa
\tau_P\pa r} -\frac{3}{2r}\sqrt{\frac{2M}{r}}\frac{\pa\Psi}{\pa \tau_P} 
-\frac{2}{r}\l[1-\frac{M}{r}\r] \frac{\pa\Psi}{\pa r} \nonumber \\ 
& & - \!\l(1-\frac{2M}{r}\r)\frac{\pa^2\Psi}{\pa r^2} 
 - \frac{l(l+1)}{r^2}\Psi = 0.
\label{painsemi1}
\er
The semiclassical wave-functions satisfying the above equation are obtained by 
making the standard ansatz for $\Psi$ {\it i.e.}, 
\b
\Psi(\tau_P,r) = \exp\left[i S(\tau_P,r)/\hbar \right]
\e
where $S$ is a function which is expanded in powers of $\hbar$ of the form
\begin{equation}
\!\!\!S(r,\tau_P)\! = S_0(r,\tau_P)\! +\! \hbar ~ S_1(r,\tau_P)\! 
+\!{\hbar}^2 S_2(r,\tau_P)\! \ldots
\label{eqn:exp}
\end{equation}
Substituting into the wave equation~(\ref{painsemi1}) and neglecting
terms of order $\hbar$ and greater, we find to the lowest
order (considering the case $l = 0$), 
\b
\!\l(1- {2M\over r}\r)\!\!\l(\frac{\pa S_0}{\pa r}\r)^2\!\!\! - \!\!
\l(\frac{\pa S_0}{\pa \tau_P}\r)^2\!\!\! + \sqrt{8M \over r}\!
\l(\frac{\pa S_0}{\pa\tau_P}\r)\!\! \l(\frac{\pa S_0}{\pa r}\r) = 0 
\label{painsemi2}
\e
The above equation is just the Hamilton-Jacobi equation satisfied by a
massless particle moving in the space-time determined by the line element 
(\ref{painleves}). The action functional($S_0$) satisfying the classical 
Hamilton-Jacobi equation will immediately give us the semiclassical Kernel 
$K(r_2,t_2;r_1,t_1)$ for the particle to propagate from 
$(t_1,r_1)$ to $(t_2, r_2)$ in the saddle point approximation. Introducing a 
dimensionless variable $\rho = r/2M$, the solution to the above equation 
is easily found to be,
\b
S_0 = -E\tau_P + 2ME \int \! d\rho {\sqrt{\rho}(1\pm \sqrt{\rho}) \over \rho
-1}
\label{painsemi3}.
\e
Choosing the positive sign in the above equation, we see that the 
solution is singular at $\rho = 1$ which corresponds to the horizon. The 
method of complex paths identifies appropriate complex contours which 
satisfies the semi-classical condition to be the one lying in the upper 
complex plane (For an extensive discussion on this, see SP). The complex 
paths, we use in 
our analysis, takes into account of all possible paths satisfying the semi-
classical ansatz - irrespective of the multiple mapping of the part/whole of 
the space-time. For the PC, as noted earlier, the whole of space-time is 
doubly mapped {\it w.r.t.} the Schwarzschild space-time implying that PC 
contains two distinct R and T regions. Hence, the semi-classical propagator 
will have contributions from complex paths from both of these regions which 
have no common point due to each path being in different R and T regions. 
Hence, the contribution to the amplitude of emission/absorption by these two 
paths will be mutually exclusive. Thus, regularizing the singularity by such 
contours and dividing the result by two to take care of the over counting of 
the paths, we obtain
\b
S_0[{\rm emission}] = {\rm real \; part} + 2i\pi ME
\label{painsemi4a}
\e
In this case, the action that has been calculated is interpreted to be that
for emission by analogy with done by SP. In order to obtain the 
action for absorption of particles, we have to consider the Hamilton-Jacobi 
equation for the upper sign in the line element (\ref{painleves}) and
repeating the above calculations and using the arguments above we obtain
\b
S_0[{\rm absorption}] = {\rm real \; part} -  2i\pi ME
\label{painsemi5}
\e
where the minus sign arises from choosing the appropriate complex contour
given by the semiclassical prescription. Taking the modulus square to obtain 
the probability, we get 
$P[{\rm emission}] \propto \exp\left(-{4 \pi M  E} \right)$ and 
$ P[{\rm absorption}] \propto \exp\left({4 \pi M E} \right)$, thus
\b
P[{\rm emission}] = \exp\left(-{8 \pi M E} 
\right)P[{\rm absorption}]. \label{eqn:ssys}
\e
The exponential dependence on the energy allows one to give a 
``thermal'' interpretation to this result, which shows that the temperature 
of the emission spectrum is the standard Hawking temperature. 
\par
SP have shown explicitly that very close to the horizon, the terms containing 
the mass and angular part does not contribute significantly. Similar results 
hold in our case and the above analysis is therefore applicable to both 
massless and massive scalar particles. 
\par
	In our interpretation, we consider the amplitude for pair creation 
both inside and outside the horizon. The semiclassical treatment of Hawking 
radiation for the PC is obtained recently by Parikh and Wilczek\cite{parikh}. 
The authors considered Hawking radiation as a pair creation outside the 
horizon, with the negative energy particle tunneling into the black hole. The 
tunneling of particles produced just inside the horizon also contributes to 
the Hawking radiation and is incorporated in our formalism.   
\par
	We now consider the quantization of massless scalar field $\Phi$ in 
the LC. The system we first consider is the lower sign in line element 
(\ref{lemaitres}). Separating the angular variables by setting $\Phi = 
\Psi(\tau, R) Y^m_l(\theta, \phi)$, making the standard semiclassical ansatz 
for $\Psi$ and expanding $S$ in powers of $\hbar$ one finds, to lowest order 
(for $l = 0$),
\b
\l(1-U^{2/3}\r)\!\l[\l(\partial_{U}S_0\r)^2\! +\! 
\l(\partial_{V}S_0\r)^2 \r]\! -\! 2\l(1+U^{2/3}\r) (\partial_{U}S_0) 
(\partial_{V}S_0) = 0 
\label{lemsemi3},
\e  
where $U$ and $V$ are dimensionless parameters given by 
\b 
U = \frac{3}{4M}\l(R - \tau\r), \qquad V = \frac{3}{4M}\l(R + \tau\r)
\label{lemsemi2}.
\e
The above equation is just the Hamilton-Jacobi equation satisfied by a 
massless particle moving in the space-time determined by the line element 
(\ref{lemaitres}). Making the ansatz $S_0 = -4 M EV/3 + f(U)$, we obtain,
\b
\frac{df}{dU} = E\frac{1+U^{2/3} \pm 2U^{1/3}}{1-U^{2/3}}
\label{lemsemi4},
\e
where the $\pm$ signs arise from taking square roots. Notice that it is only
for the positive sign that the denominator is singular at $U=1$. This singular
solution for $f$ evidently corresponds to outgoing particles. Therefore
choosing the positive sign and making the convenient change of variable $x^3 =
U$ the solution to Eq.~(\ref{lemsemi3}) is given by
\b
S_0 = -\frac{4ME}{3} V + 4M E\int\! dx \frac{x^2(1+x)}{1-x}
\label{lemsemi5}.
\e
It is clear that the action function is again singular at the horizon $x=1$. 
Using the method of complex paths, we regularize the singularity by a complex 
contour lying in the upper complex plane and obtain
\b
S_0[{\rm emission}] = {\rm real \; part} +  8i\pi ME
\label{lemsemi7}
\e
To obtain the action for absorption of particles, we repeat the above 
calculation for the upper sign in line element (\ref{lemaitres}). In this 
case, it is easy to see that the only singular solution corresponds to 
in-going particles and using the method of complex paths as in previous case, 
we obtain 
\b
S_0[{\rm absorption}] = {\rm real \; part} -  8i\pi ME.
\label{lemsemi8}
\e
As seen earlier, the complex paths takes into account of all possible paths 
satisfying the semi-classical ansatz - irrespective of the multiple mapping of 
the space-time. For the PC, the contribution to the amplitude for absorption/
emission by the complex paths are mutually exclusive as the 
paths are in two distinct R and T regions. In the case of LC, however, only 
part of the region(T) is doubly mapped, it is always possible to find one 
point that is common to the paths contributing to absorption/emission. Hence, 
these paths are not mutually exclusive in calculating amplitudes. These paths, 
on the other hand, will be mutually exclusive when one considers the 
probability amplitude. Hence, the probability amplitude for absorption/
emission obtained by taking the modulus square of the semi-classical wave 
function will have equal contributions from these two paths. Thus, the total 
contribution to the probability is from four similar paths. Taking this into 
account, by dividing $S_0$ by four, we obtain
\b
P[{\rm emission}] = \exp\left(-{8 \pi ME}\right)P[{\rm absorption}].
\e
The exponential dependence of the energy allows us to give the thermal 
interpretation to this result, which shows that the temperature of the 
emission spectrum is the standard Hawking temperature.   
\par
	It is normally assumed that the evaporation process results 
from an instability of the vacuum in the presence of the background metric. 
The particles are produced at a constant rate suggesting that the Hawking 
radiation converts the mass of the black hole into energy, thereby decreasing 
the mass. The decrease in the black hole mass is a physical effect and should 
be independent of the coordinate transformations and hence {\it Hawking 
radiation should be covariant}. Here we have shown that this is indeed true by 
the method of complex paths. 
\par
The above results can be summarized as follows: The analysis of the space-time 
structure of the two coordinates, LC and PC, using R and T regions provides an 
elegant method in understanding the global structure of the space-time, thus 
making the detailed analysis of particle trajectories unnecessary. The complex 
paths takes into account all 
paths irrespective of the multiple mapping of the part/whole of the space-
time. This is crucial to obtain the correct temperature associated with the 
Hawking radiation in these two coordinates. It has been 
shown by Davies\cite{davies76} that a freely falling detector will see a 
particle spectrum different from the thermal spectrum. This shows that 
there is no correspondence between the particles detected by the model 
detector and the particle spectrum obtained by the field theoretic analysis -- 
a result known in other contexts as well (see, for example Sriramkumar and
Padmanabhan\cite{schrambo98} and references therein).
\par
The authors would wish to thank Prof. M.A.H. MacCallum for providing with 
the English translation of Novikov's work before it's publication in GRG.  S.S. 
is being supported by the Council of Scientific and Industrial Research,
India.

\nonumsection{References}

\end{document}